\documentclass[12pt,a4paper,english,nofootinbib]{revtex4}
\usepackage{lmodern}
\usepackage{lmodern}

\usepackage[T1]{fontenc}
\usepackage[latin9]{inputenc}
\usepackage{babel}
\usepackage{mathrsfs}
\usepackage{amsmath}
\usepackage{amssymb}
\usepackage{esint}
\usepackage[unicode=true,pdfusetitle,
 bookmarks=true,bookmarksnumbered=false,bookmarksopen=false,
 breaklinks=false,pdfborder={0 0 1},backref=false,colorlinks=false]
 {hyperref}

\makeatletter

\pdfpageheight\paperheight
\pdfpagewidth\paperwidth

\@ifundefined{textcolor}{}
{%
 \definecolor{BLACK}{gray}{0}
 \definecolor{WHITE}{gray}{1}
 \definecolor{RED}{rgb}{1,0,0}
 \definecolor{GREEN}{rgb}{0,1,0}
 \definecolor{BLUE}{rgb}{0,0,1}
 \definecolor{CYAN}{cmyk}{1,0,0,0}
 \definecolor{MAGENTA}{cmyk}{0,1,0,0}
 \definecolor{YELLOW}{cmyk}{0,0,1,0}
}

\usepackage{babel}
\usepackage{babel}
\usepackage{babel}
\usepackage{babel}
\usepackage{babel}
\usepackage{babel}
\usepackage{babel}
\usepackage{babel}

\usepackage{babel}
\usepackage{graphicx}
\def\b{\begin{equation}}
\def\e{\end{equation}}

\@ifundefined{textcolor}{}{%
 \definecolor{BLACK}{gray}{0}
 \definecolor{WHITE}{gray}{1}
 \definecolor{RED}{rgb}{1,0,0}
 \definecolor{GREEN}{rgb}{0,1,0}
 \definecolor{BLUE}{rgb}{0,0,1}
 \definecolor{CYAN}{cmyk}{1,0,0,0}
 \definecolor{MAGENTA}{cmyk}{0,1,0,0}
 \definecolor{YELLOW}{cmyk}{0,0,1,0}
 }

\usepackage{latexsym}\usepackage{bm}

\makeatother

\begin{document}
\title{{{}{}{}{}Second Order Perturbation Theory in
General Relativity: Taub Charges as Integral Constraints }}
\author{{\normalsize{}{}{}{}Emel Altas}}
\email{emelaltas@kmu.edu.tr}

\affiliation{Department of Physics,\\
 Karamanoglu Mehmetbey University, 70100, Karaman, Turkey}
\author{{\normalsize{}{}{}{}Bayram Tekin}}
\email{btekin@metu.edu.tr}

\affiliation{Department of Physics,\\
 Middle East Technical University, 06800, Ankara, Turkey}
\date{{{}{}{}{}\today}}
\begin{abstract}
In a nonlinear theory, such as General Relativity, linearized field
equations around an exact solution are necessary but not sufficient conditions for linearized
solutions. Therefore, the linearized field equations can have some solutions which do not come from the linearization
of possible exact solutions. This fact can make the perturbation theory
ill-defined, which would be a problem both at the classical and semiclassical quantization
level. Here we study the first and second order perturbation theory
in cosmological Einstein gravity and give the explicit form of the
integral constraint, which is called the Taub charge, on the first
order solutions for spacetimes with a Killing symmetry and a compact
hypersurface without a boundary. 
\end{abstract}
\maketitle

\section{{\normalsize{}{}{}{}Introduction}}

Let us consider a generic gravity theory defined (in a vacuum) by
the nonlinear field equations 
\begin{equation}
\mathscr{E}_{\mu\nu}(g)=0,\label{genericfieldequation}
\end{equation}
in some local coordinates. We assume the usual "Bianchi Identity",
$\nabla^{\mu}\mathscr{E}_{\mu\nu}=0$ which plays a central role in the ensuing discussion. The physical situation (the spacetime)  as an
exact solution is often too difficult to construct. Hence one resorts
to perturbation theory around a symmetric background solution $\bar{g}$,
and expands (\ref{genericfieldequation}) as 
\begin{equation}
\bar{\mathscr{E}}_{\mu\nu}(\bar{g})+\lambda(\mathscr{E}_{\mu\nu})^{\left(1\right)}(h)+\lambda^{2}\bigg((\mathscr{E}_{\mu\nu})^{\left(2\right)}(h,h)+(\mathscr{E}_{\mu\nu})^{\left(1\right)}(k)\bigg)+ {\mathcal{O}}(\lambda^3)=0,\label{expansion1}
\end{equation}
where $\lambda$ is a dimensionless small parameter introduced to
keep track of the formal perturbative expansion; and the $h$ and
$k$ tensor fields are defined as 
\begin{equation}
h_{\mu\nu}:={\frac{d}{d\lambda}g_{\mu\nu}}\bigg\rvert_{\lambda=0},~~~~~~~~~~~~~k_{\mu\nu}:=\frac{d^{2}}{d\lambda^{2}}g_{\mu\nu}\bigg\rvert_{\lambda=0}. \label{def}
\end{equation}
So as the notation suggests: $(\mathscr{E}_{\mu\nu})^{\left(1\right)}(h)$ is the linearization of the $\mathscr{E}_{\mu\nu}$ coming from the expansion of $\mathscr{E}_{\mu\nu}( \bar {g}+ \lambda h + \lambda^2 k)$, while the second order terms come in two different form as shown in (\ref{expansion1}).
At the lowest order, one sets $\bar{\mathscr{E}}_{\mu\nu}(\bar{g})=0$ and at
the first order the linearized field equations read 
\begin{equation}
(\mathscr{E}_{\mu\nu})^{\left(1\right)}(h)=0.
\end{equation}
It is clear that these equations are a {\it necessary} conditions on the 
first order perturbation $h$ defined via (\ref{def}). But, the crucial point is the following:
generically not all solutions of the linearized equations are
viable solutions since from (\ref{expansion1}), at the second order
we have the equation: 
\begin{equation}
(\mathscr{E}_{\mu\nu})^{\left(2\right)}(h,h)+(\mathscr{E}_{\mu\nu})^{\left(1\right)}(k)=0.\label{secondorderperturbedfieldequations}
\end{equation}
Upon a cursory look, this equation basically says that $(\mathscr{E}_{\mu\nu})^{\left(2\right)}(h,h)$
is a ''source'' for the second order perturbation $k$. Thus, in principle
whenever the operator $(\mathscr{E}_{\mu\nu})^{\left(1\right)}(.)$
is invertible one has a solution. Typically, due to gauge invariance
$(\mathscr{E}_{\mu\nu})^{\left(1\right)}(.)$ is not invertible but
after gauge fixing, it can be made invertible. This is a well-known, but easily remediable problem either with some locally or globally valid gauges, such as the de Donder gauge. So this is not the issue that we are interested in here. Even if a proper gauge is found, there are still situations where (\ref{secondorderperturbedfieldequations})
leads to constraints on the {\it first order perturbation} $h$ for a non-trivial solution $k$. As the basic premise of perturbation theory is its improvability by adding more terms, generically $k$ has to exist without a need to modify the first order perturbation $h$; stated in another way $h$ must be an integrable deformation. 

To see the constraints, let
$\bar{\xi}^{\mu}$ be a Killing vector field of the background metric
$\bar{g}_{\mu\nu}$. Then contracting equation (\ref{secondorderperturbedfieldequations})
with $\bar{\xi}^{\mu}$ and integrating over a hypersurface $\varSigma$
of the spacetime manifold $\mathscr{M}$, one has the constraint 
\begin{equation}
\intop_{\varSigma}d^{n-1}x\thinspace\sqrt{\bar{\gamma}}\thinspace\bar{\xi}_{\mu}\thinspace(\mathscr{E}^{\mu\nu})^{\left(1\right)}(k)=-\intop_{\varSigma}d^{n-1}x\thinspace\sqrt{\bar{\gamma}}\thinspace\bar{\xi}_{\mu}\thinspace(\mathscr{E}^{\mu\nu})^{\left(2\right)}(h,h), \label{bye}
\end{equation}
where one uses the background metric and its inverse to lower and
raise the indices. $\bar{\gamma}$ is the metric on the hypersurface.
The left-hand side can be written as a boundary term as 
\begin{equation}
\sqrt{\bar{\gamma}}\bar{\xi}_{\mu}\thinspace(\mathscr{E}^{\mu\nu})^{\left(1\right)}(k)=\partial_{\mu}\left(\sqrt{\bar{\gamma}}F^{\mu\nu}(\bar{\xi},k)\right),
\end{equation}
where $F^{\mu\nu}(\bar{\xi},k)$ is an antisymmetric tensor
field. For more details on this see \cite{altasuzunmakale,emeltez}.
The left-hand side of (\ref{bye}), when $h$ is used, is called the Abbott-Deser-Tekin (ADT) \cite{AD,DT} (an extension of the ADM \cite{ADM}) 
and the right-hand side of (\ref{bye}) is called the Taub charge \cite{Taub}. So
we have the equality of the ADT and Taub charges as a constraint at
the second order in perturbation theory for the case when the background spacetime has at least one Killing vector field:
\begin{equation}
Q_{ADT}\left[\bar{\xi}\right]:=\intop_{\partial\varSigma}d\varSigma_{\mu}\thinspace\sqrt{\bar{\sigma}}\thinspace\hat{n}_{\nu}\thinspace\bar{\xi}_{\mu}\thinspace F^{\mu\nu}(\bar{\xi},k)=-\intop_{\varSigma}d^{n-1}x\thinspace\sqrt{\bar{\gamma}}\thinspace\bar{\xi}_{\mu}\thinspace(\mathscr{E}^{\mu\nu})^{\left(2\right)}(h,h)=:-Q_{Taub}\left[\bar{\xi}\right],\label{equalityofcharges}
\end{equation}
where $\bar{\sigma}$ is the metric on $\partial\varSigma$ and $\hat{n}_{\nu}$
is the outward unit normal vector on it. If $\varSigma$ does not
have a boundary, then the ADT charges vanish identically and so must
the Taub charges. The vanishing of the Taub charges is not automatic, therefore, one has an apparent integral constraint
on the linearized solution $h$ as: 
\begin{equation}
\intop_{\varSigma}d^{n-1}x\thinspace\sqrt{\bar{\gamma}}\thinspace\bar{\xi}_{\mu}\thinspace(\mathscr{E}^{\mu\nu})^{\left(2\right)}(h,h)=0\label{integralconstraint}
\end{equation}
on a compact surface without a boundary. If this constraint were to be
satisfied, then $h$ would be a generic linearized solution which can be
added to $\bar{g}$ to improve the exact solution. On the other hand,
if (\ref{integralconstraint}) is not satisfied, then
one speaks of a linearization instability. This issue was studied
in various aspects in \cite{Deser-Brill,Deser-Bruhat,Fischer-Marsden,Fischer-Marsden-Moncrief,Marsden,Moncrief,Arms-Marsden}
for Einstein's theory and summarized in \cite{Bruhat,Girbau-Bruna};
and extended to generic gravity theories more recently \cite{altasuzunmakale,emeltez, Altas_chiral}.
From these works two main conclusions follow: first, in Einstein's
theory, a solution set to the constrained initial data on a compact Cauchy
surface without a boundary may not have nearby solutions, hence they
can be isolated and perturbations are not allowed; second, for generic
gravity theories in asymptotically (anti) de Sitter spacetimes,
linearization instability arises for certain combinations of the parameters
defining the theory. 

Regarding (\ref{integralconstraint}), the obvious
question is whether $\sqrt{\bar{\gamma}}\thinspace\bar{\xi}_{\mu}\thinspace(\mathscr{E}^{\mu\nu})^{\left(2\right)}(h,h)$
is a boundary term for Einstein's gravity or not: if it were a boundary term, one would
not have the linearization instability observed in the previous works,
because it would also vanish identically on a manifold without a boundary.
Here for cosmological Einstein's theory we show explicitly that $\bar{\xi}_{\mu}\thinspace(\mathscr{E}^{\mu\nu})^{\left(2\right)}(h,h)$
has a bulk and a boundary part, the later drops for the case of compact
hypersurface without a boundary while the former is a constraint on
the first order perturbation.

The lay out of the paper is as follows: in section $\text{\mbox{II}}$
we give the details of the first order expression for the cosmological
Einstein tensor in a generic Einstein spacetime in terms of the perturbation
$h$ and give a concise formula in terms of the linearized Riemann
tensor for (anti) de Sitter backgrounds using our results \cite{newformula,newformulauzun}.
In section $\text{\mbox{III}}$ we study the second order cosmological
Einstein tensor in a generic Einstein background and specify to the
case of (anti) de Sitter. In section $\text{\mbox{IV}}$ we discuss
the gauge invariance issue and relegate some of the computations to
the Appendices.

\section{First order perturbation theory}

Here to set the stage, we recapitulate what is already known in the first order perturbation theory in a generic Einstein background. Using the results of Appendix A, one can show that the linearized cosmological Einstein tensor about a generic Einstein space, defined as\footnote{We shall denote $(\ensuremath{{\cal {G}}}_{\mu\nu}){}^{\left(1\right)}(h)$
as $(\ensuremath{{\cal {G}}}_{\mu\nu}){}^{\left(1\right)}$ and $(\ensuremath{{\cal {G}}}_{\mu\nu}){}^{\left(2\right)}(h,h)$
as $(\ensuremath{{\cal {G}}}_{\mu\nu}){}^{\left(2\right)}$ and we
use the some notation for other tensors. }
\begin{equation}
(\ensuremath{{\cal {G}}}_{\mu\nu}){}^{\left(1\right)}:=(R_{\mu\nu}){}^{\left(1\right)}-\frac{1}{2}\bar{g}_{\mu\nu}(R){}^{\left(1\right)}-\frac{1}{2}h_{\mu\nu}\bar{R}+\Lambda h_{\mu\nu}
\end{equation}
can be written as a divergence plus a residual part \cite{DT,sisman-tekin-setare}
\begin{equation}
\left({\cal G}^{\mu\nu}\right)^{\left(1\right)}=\bar{\nabla}_{\alpha}\bar{\nabla}_{\beta}K^{\mu\alpha\nu\beta}+X^{\mu\nu},
\end{equation}
where  the $K$-tensor reads 
\begin{equation}
K^{\mu\alpha\nu\beta}\equiv\frac{1}{2}\left(\bar{g}^{\alpha\nu}\tilde{h}^{\mu\beta}+\bar{g}^{\mu\beta}\tilde{h}^{\alpha\nu}-\bar{g}^{\alpha\beta}\tilde{h}^{\mu\nu}-\bar{g}^{\mu\nu}\tilde{h}^{\alpha\beta}\right),\qquad\tilde{h}^{\mu\nu}\equiv h^{\mu\nu}-\frac{1}{2}\bar{g}^{\mu\nu}h,
\end{equation}
and the residual tensor reads 
\begin{align}
X^{\mu\nu}  \equiv
  \frac{1}{2}\left(h^{\mu\alpha}\bar{R}_{\alpha}\thinspace^{\nu}-\bar{R}^{\mu\alpha\nu\beta}h_{\alpha\beta}\right)+\frac{1}{2}\bar{g}^{\mu\nu}h^{\rho\sigma}\bar{R}_{\rho\sigma}+\Lambda h^{\mu\nu}-\frac{1}{2}h^{\mu\nu}\bar{R}.
\end{align}
The background conserved current can be obtained via contracting the linearized cosmological Einstein tensor with the background Killing vector $\bar{\xi}_{\nu}$ to get
\begin{equation}
\bar{\xi}_{\nu}\left({\cal G}^{\mu\nu}\right)^{\left(1\right)}=\bar{\nabla}_{\alpha}\left(\bar{\xi}_{\nu}\bar{\nabla}_{\beta}K^{\mu\alpha\nu\beta}-K^{\mu\beta\nu\alpha}\bar{\nabla}_{\beta}\bar{\xi}_{\nu}\right)+K^{\mu\alpha\nu\beta}\bar{R}_{\thinspace\beta\alpha\nu}^{\rho}\bar{\xi}_{\rho}+X^{\mu\nu}\bar{\xi}_{\nu}.
\label{onemli}
\end{equation}
The non-divergence terms cancel upon use of the field equations and therefore one has a pure boundary term 
\begin{equation}
\bar{\xi}_{\nu}\left({\cal G}^{\mu\nu}\right)^{\left(1\right)}=\bar{\nabla}_{\alpha}F^{\alpha\mu}\left(\bar{\xi},h\right)
\end{equation}
with 
\begin{equation}
F^{\alpha\mu}\left(\bar{\xi},h\right)=\bar{\xi}_{\nu}\bar{\nabla}_{\beta}K^{\mu\alpha\nu\beta}-K^{\mu\beta\nu\alpha}\bar{\nabla}_{\beta}\bar{\xi}_{\nu}.\label{boundaryAD}
\end{equation}
It is important to note that $\left({\cal G}^{\mu\nu}\right)^{\left(1\right)}$ is a background gauge invariant tensor, hence the above expression is gauge invariant;
but $F^{\mu\nu}\left(\bar{\xi},h\right)$ itself is
only gauge invariant up to a boundary term whose divergence vanishes. The above result is valid for generic Einstein backgrounds.  For (anti) de Sitter spacetimes, one can do better and express $F^{\mu\nu}\left(\bar{\xi}, h\right)$ in an exactly gauge invariant way \cite{newformula,newformulauzun}. For this purpose, let us introduce a new tensor, which we called the $\ensuremath{{\cal {P}}}$-tensor, as 
\begin{equation}
\text{\ensuremath{{\cal {P}}}}^{\nu\mu}\thinspace_{\beta\sigma}:=R^{\nu\mu}\thinspace_{\beta\sigma}+\delta_{\sigma}^{\nu}\text{\ensuremath{{\cal {G}}}}_{\beta}^{\mu}-\delta_{\beta}^{\nu}\text{\ensuremath{{\cal {G}}}}_{\sigma}^{\mu}+\delta_{\beta}^{\mu}\text{\ensuremath{{\cal {G}}}}_{\sigma}^{\nu}-\delta_{\sigma}^{\mu}\text{\ensuremath{{\cal {G}}}}_{\beta}^{\nu}+\left(\frac{R}{2}-\frac{\Lambda\left(n+1\right)}{n-1}\right)\left(\delta_{\sigma}^{\nu}\delta_{\beta}^{\mu}-\delta_{\beta}^{\nu}\delta_{\sigma}^{\mu}\right),\label{eq:ptensoruudd-1}
\end{equation}
which has the following nice properties: 
\begin{itemize}
\item It has the symmetries of the Riemann tensor. 
\item It is divergence free, $\nabla_{\nu}\text{\ensuremath{{\cal {P}}}}^{\nu\mu}\thinspace_{\beta\sigma}=0$. 
\item Its trace is the cosmological Einstein tensor, $\text{\ensuremath{{\cal {P}}}}^{\mu}\thinspace_{\sigma}:=\text{\ensuremath{{\cal {P}}}}^{\nu\mu}\thinspace_{\nu\sigma}=(3-n)\text{\ensuremath{{\cal {G}}}}_{\sigma}^{\mu}$. 
\item When evaluated for a background Einstein space, it yields 
\begin{equation*}
\bar{\text{\ensuremath{{\cal {P}}}}}^{\nu\mu}\thinspace_{\beta\sigma}=\bar{R}^{\nu\mu}\thinspace_{\beta\sigma}+\frac{2\Lambda}{(n-1)(n-2)}\left(\delta_{\sigma}^{\nu}\delta_{\beta}^{\mu}-\delta_{\beta}^{\nu}\delta_{\sigma}^{\mu}\right),\label{backgroundptensor}
\end{equation*}
and so 
\begin{equation*}
\bar{\text{\ensuremath{{\cal {P}}}}}^{\nu\mu}\thinspace_{\beta\sigma}=\bar{C}^{\nu\mu}\thinspace_{\beta\sigma},
\end{equation*}
where $C^{\nu\mu}\thinspace_{\beta\sigma}$ is the Weyl tensor which
vanishes for (anti) de Sitter spacetimes.
\end{itemize}
From the above construction, it is clear that the formalism works for $n \ge 4$ dimensions; therefore we shall assume this in the ensuing discussion.  Using all these properties, one can show that at first order the covariantly
conserved current is a total derivative
\begin{equation}
\bar{\text{\ensuremath{\xi}}}_{\nu}(\text{\ensuremath{{\cal {G}}}}^{\nu\mu})^{\left(1\right)}=\frac{(n-1)(n-2)}{4\Lambda\left(n-3\right)}\bar{\nabla_{\nu}}\biggl((\text{\ensuremath{{\cal {P}}}}^{\nu\mu}\thinspace_{\beta\sigma})^{\left(1\right)}\bar{\nabla}^{\beta}\bar{\xi}^{\sigma}\biggr),\label{eq:finallinearequation}
\end{equation}
where the first order linearization of the $\text{\ensuremath{{\cal {P}}}}$-tensor
in (anti) de Sitter spacetime reads
\begin{equation}
({\cal {P}}^{\nu\mu}\thinspace_{\beta\sigma})^{\left(1\right)}=(R^{\nu\mu}\thinspace_{\beta\sigma})^{(1)}+2(\text{\ensuremath{{\cal {G}}}}_{[\beta}^{\mu})^{(1)}\delta_{\sigma]}^{\nu}+2(\text{\ensuremath{{\cal {G}}}}_{[\sigma}^{\nu})^{(1)}\delta_{\beta]}^{\mu}+(R)^{\left(1\right)}\delta_{[\beta}^{\mu}\delta_{\sigma]}^{\nu}.\label{ktensorlinear}
\end{equation}
Making use of this construction one has the conserved charge in a
compact form:
\begin{equation}
\boxed{\phantom{\frac{\frac{\xi}{\xi}}{\frac{\xi}{\xi}}}Q\left[\bar{\xi}\right]=\frac{(n-1)(n-2)}{8(n-3)\Lambda G\Omega_{n-2}}\int_{\partial\bar{\Sigma}}d^{n-2}x\,\sqrt{\bar{\sigma}}\,\bar{n}_{\mu}\bar{\sigma}_{\nu}\left(R^{\nu\mu}\thinspace_{\beta\sigma}\right)^{\left(1\right)}\bar{\nabla}^{\beta}\bar{\xi}^{\sigma},\phantom{\frac{\frac{\xi}{\xi}}{\frac{\xi}{\xi}}}}\label{newcharge}
\end{equation}
where we used the fact that 
$(\text{\ensuremath{{\cal {G}}}}^{\mu\nu})^{\left(1\right)}=0$
and $(R)^{(1)}=0$ on the boundary. Here $\bar{\sigma}_\nu$ is the unit outward normal vector on $\partial\bar{\Sigma}$. Gauge transformation properties
are discussed below in Section $\text{\mbox{IV}}$ in more detail.
But here, let us note that under a variation generated by the vector
field $X$, which we denote as $\delta_{X}$, one has $\delta_{X}\left(R^{\nu\mu}\thinspace_{\beta\sigma}\right)^{\left(1\right)}=\mathscr{L}_{X}\bar{R}^{\nu\mu}\thinspace_{\beta\sigma}$
which vanishes for (anti) de Sitter backgrounds (see section %
\mbox{%
III%
} of \cite{newformulauzun} for more details and for the gauge transformation
properties of the expression \ref{boundaryAD}). Let us now turn to our main goal of computing
the analogous expression at second order. 

\section{Second order perturbation theory }

For any antisymmetric two tensor $\text{\ensuremath{{\cal {F}}}}^{\beta\sigma}$,
one has the exact identity
\begin{equation}
\nabla_{\nu}(\text{\ensuremath{{\cal {F}}}}^{\beta\sigma}\text{\ensuremath{{\cal {P}}}}^{\nu\mu}\thinspace_{\beta\sigma})-\text{\ensuremath{{\cal {P}}}}^{\nu\mu}\thinspace_{\beta\sigma}\nabla_{\nu}\text{\ensuremath{{\cal {F}}}}^{\beta\sigma}=0.\label{eq:mainequationuudd}
\end{equation}
Soon we will choose $\text{\ensuremath{{\cal {F}}}}^{\beta\sigma}$ to be the {\it potential } of the Killing vector field below. Expansion of this identity at second order yields
\begin{eqnarray}
 &  & \bar{\nabla}_{\nu}\bigg(\bar{\text{\ensuremath{{\cal {F}}}}}^{\beta\sigma}(\text{\ensuremath{{\cal {P}}}}^{\nu\mu}\thinspace_{\beta\sigma})^{\left(2\right)}+(\text{\ensuremath{{\cal {F}}}}^{\beta\sigma})^{\left(1\right)}(\text{\ensuremath{{\cal {P}}}}^{\nu\mu}\thinspace_{\beta\sigma})^{\left(1\right)}+(\text{\ensuremath{{\cal {F}}}}^{\beta\sigma})^{\left(2\right)}\bar{\text{\ensuremath{{\cal {P}}}}}^{\nu\mu}\thinspace_{\beta\sigma}\bigg)-2(\Gamma_{\nu\rho}^{\beta})^{\left(1\right)}\bar{\text{\ensuremath{{\cal {F}}}}}^{\rho\sigma}(\text{\ensuremath{{\cal {P}}}}^{\nu\mu}\thinspace_{\beta\sigma})^{\left(1\right)}\nonumber \\
 &  & -(\text{\ensuremath{{\cal {P}}}}^{\nu\mu}\thinspace_{\beta\sigma})^{\left(1\right)}\bar{\nabla}_{\nu}(\text{\ensuremath{{\cal {F}}}}^{\beta\sigma})^{\left(1\right)}-\bar{\text{\ensuremath{{\cal {P}}}}}^{\nu\mu}\thinspace_{\beta\sigma}\bar{\nabla}_{\nu}(\text{\ensuremath{{\cal {F}}}}^{\beta\sigma})^{\left(2\right)}-(\text{\ensuremath{{\cal {P}}}}^{\nu\mu}\thinspace_{\beta\sigma})^{\left(2\right)}\bar{\nabla}_{\nu}\bar{\text{\ensuremath{{\cal {F}}}}}^{\beta\sigma}+(\Gamma_{\nu\rho}^{\nu})^{\left(2\right)}\bar{\text{\ensuremath{{\cal {F}}}}}^{\beta\sigma}\bar{\text{\ensuremath{{\cal {P}}}}}^{\rho\mu}\thinspace_{\beta\sigma}\nonumber \\
 &  & +(\Gamma_{\nu\rho}^{\nu})^{\left(1\right)}\bigg(\bar{\text{\ensuremath{{\cal {F}}}}}^{\beta\sigma}(\text{\ensuremath{{\cal {P}}}}^{\rho\mu}\thinspace_{\beta\sigma})^{\left(1\right)}+(\text{\ensuremath{{\cal {F}}}}^{\beta\sigma})^{\left(1\right)}\bar{\text{\ensuremath{{\cal {P}}}}}^{\rho\mu}\thinspace_{\beta\sigma}\bigg)-2(\Gamma_{\nu\rho}^{\beta})^{\left(2\right)}\bar{\text{\ensuremath{{\cal {F}}}}}^{\rho\sigma}\bar{\text{\ensuremath{{\cal {P}}}}}^{\nu\mu}\thinspace_{\beta\sigma}\nonumber \\
 &  & -2(\Gamma_{\nu\rho}^{\beta})^{\left(1\right)}\bigg((\text{\ensuremath{{\cal {F}}}}^{\rho\sigma})^{\left(1\right)}\bar{\text{\ensuremath{{\cal {P}}}}}^{\nu\mu}\thinspace_{\beta\sigma}+\bar{\text{\ensuremath{{\cal {F}}}}}^{\rho\sigma}(\text{\ensuremath{{\cal {P}}}}^{\nu\mu}\thinspace_{\beta\sigma})^{\left(1\right)}\bigg)=0.\label{secondorderperturbedeq}
\end{eqnarray}
Making use of the first order linearization of the Bianchi-type identity $\nabla_{\nu}\text{\ensuremath{{\cal {P}}}}{}^{\nu\mu}\thinspace_{\beta\sigma}=0$,
that is 
\begin{equation}
\bar{\nabla}_{\nu}(\text{\ensuremath{{\cal {P}}}}{}^{\nu\mu}\thinspace_{\beta\sigma})^{\left(1\right)}-(\Gamma_{\nu\beta}^{\rho})^{\left(1\right)}\bar{\text{\ensuremath{{\cal {P}}}}}^{\nu\mu}\thinspace_{\rho\sigma}-(\Gamma_{\nu\sigma}^{\rho})^{\left(1\right)}\bar{\text{\ensuremath{{\cal {P}}}}}^{\nu\mu}\thinspace_{\beta\rho}+(\Gamma_{\nu\rho}^{\nu})^{\left(1\right)}\bar{\text{\ensuremath{{\cal {P}}}}}^{\rho\mu}\thinspace_{\beta\sigma}=0,\label{linearorderdivergenceptensor}
\end{equation}
and taking $\bar{{\cal {F}}}^{\rho\sigma} = \bar{\nabla}^\rho \bar{\xi}^\sigma$,
(\ref{secondorderperturbedeq}) reduces to 
 \begin{eqnarray}
 &  & \bar{\nabla}_{\nu}\left(\bar{\text{\ensuremath{{\cal {F}}}}}^{\beta\sigma}(T^{\nu\mu}\thinspace_{\beta\sigma})^{\left(2\right)}\right)-\bar{R}_{\lambda\nu}\thinspace^{\beta\sigma}\bar{\xi}^{\lambda}(T^{\nu\mu}\thinspace_{\beta\sigma})^{\left(2\right)}-2(\Gamma_{\nu\rho}^{\beta})^{\left(1\right)}\bar{\text{\ensuremath{{\cal {F}}}}}^{\rho\sigma}(\text{\ensuremath{{\cal {P}}}}^{\nu\mu}\thinspace_{\beta\sigma})^{\left(1\right)}\nonumber \\
 &  & +\bar{\text{\ensuremath{{\cal {P}}}}}^{\nu\mu}\thinspace_{\beta\sigma}\bar{\text{\ensuremath{{\cal {F}}}}}^{\gamma\sigma}(\delta_{\gamma}^{\beta}\delta_{\lambda}^{\rho}-2\delta_{\lambda}^{\beta}\delta_{\gamma}^{\rho})(\Gamma_{\nu\rho}^{\delta})^{\left(1\right)}\left(\frac{h}{2}\delta_{\delta}^{\lambda}-h_{\delta}^{\lambda}\right)=0,\label{rereduced}
\end{eqnarray}
where for notational simplicity, we introduced
\begin{equation}
(T^{\nu\mu}\thinspace_{\beta\sigma})^{\left(2\right)}:=(\text{\ensuremath{{\cal {P}}}}^{\nu\mu}\thinspace_{\beta\sigma})^{\left(2\right)}+\frac{h}{2}(\text{\ensuremath{{\cal {P}}}}^{\nu\mu}\thinspace_{\beta\sigma})^{\left(1\right)}.
\end{equation}
Rewriting the algebraic decomposition of the Riemann tensor, one finds final expression in terms of the background Weyl tensor as
\begin{eqnarray}
\bar{\xi}^{\nu}(\ensuremath{{\cal {G}}}_{\nu}^{\mu})^{\left(2\right)}=\frac{(n-1)(n-2)}{4\Lambda(n-3)}\Biggl(\bar{\nabla}_{\nu}\Bigl(\bar{\text{\ensuremath{{\cal {F}}}}}^{\beta\sigma}(T^{\nu\mu}\thinspace_{\beta\sigma})^{\left(2\right)}\Bigr)-2(\Gamma_{\nu\rho}^{\beta})^{\left(1\right)}\bar{\text{\ensuremath{{\cal {F}}}}}^{\rho\sigma}(\text{\ensuremath{{\cal {P}}}}^{\nu\mu}\thinspace_{\beta\sigma})^{\left(1\right)}\nonumber \\
-\bar{C}_{\lambda\nu}\thinspace^{\beta\sigma}\bar{\xi}^{\lambda}(T^{\nu\mu}\thinspace_{\beta\sigma})^{\left(2\right)}+\bar{C}^{\nu\mu}\thinspace_{\beta\sigma}\bar{\text{\ensuremath{{\cal {F}}}}}^{\gamma\sigma}(\delta_{\gamma}^{\beta}\delta_{\lambda}^{\rho}-2\delta_{\lambda}^{\beta}\delta_{\gamma}^{\rho})(\Gamma_{\nu\rho}^{\delta})^{\left(1\right)}\left(\frac{h}{2}\delta_{\delta}^{\lambda}-h_{\delta}^{\lambda}\right)\Biggr).\label{backgroundeinstein}
\end{eqnarray}
This is still a rather complicated expression having a divergence part and non-divergent parts. What we know is that,
one has $\bar{\nabla}_{\mu}(\bar{\xi}^{\nu}(\ensuremath{{\cal {G}}}_{\nu}^{\mu})^{\left(2\right)})=0.$
The main question was to show that $\bar{\xi}^{\nu}(\ensuremath{{\cal {G}}}_{\nu}^{\mu})^{\left(2\right)}$
is not a pure divergence. One can try to simplify (\ref{backgroundeinstein})
further to recast it in a pure divergence form, but there always remain
some terms outside the derivative. One can work out the details in
the more manageable (anti) de Sitter case for which the Weyl tensor vanishes;
and one ends up with 
\begin{equation}
\boxed{\phantom{\frac{\frac{\xi}{\xi}}{\frac{\xi}{\xi}}}\bar{\xi}^{\nu}(\ensuremath{{\cal {G}}}_{\nu}^{\mu})^{\left(2\right)}=\frac{(n-1)(n-2)}{4\Lambda(n-3)}\Biggl(\bar{\nabla}_{\nu}\Bigl(\bar{\text{\ensuremath{{\cal {F}}}}}^{\beta\sigma}(T^{\nu\mu}\thinspace_{\beta\sigma})^{\left(2\right)}\Bigr)-2(\Gamma_{\nu\rho}^{\beta})^{\left(1\right)}\bar{\text{\ensuremath{{\cal {F}}}}}^{\rho\sigma}(\text{\ensuremath{{\cal {P}}}}^{\nu\mu}\thinspace_{\beta\sigma})^{\left(1\right)}\Biggr).\phantom{\frac{\frac{\xi}{\xi}}{\frac{\xi}{\xi}}}}\label{secondordermainequation}
\end{equation}
From this expression and from (\ref{integralconstraint}) one finds
that on a manifold with a compact hypersurface $\Sigma$ without a
boundary, all the first order solutions $h_{\mu\nu}$ of $(\ensuremath{{\cal {G}}}^{\mu\nu})^{\left(1\right)}=0$,
must also satisfy the second order integral constraint for $n>3$
\begin{equation}
\frac{1}{\Lambda}\intop_{\Sigma}d^{n-1}x\thinspace\sqrt{\bar{\gamma}}\thinspace(\Gamma_{\nu\rho}^{\beta})^{\left(1\right)}\bar{\nabla}^\rho \bar{\xi}^\sigma(R^{\nu\mu}\thinspace_{\beta\sigma})^{\left(1\right)}=0.
\end{equation}
Any first order solution that does not satisfy this automatically
cannot come from the linearization of an exact metric. Stated in a more
geometric vantage point, such solutions do not lie in the tangent
space of the "point" $\bar{g}$ in the space of solutions, they are artifacts of linearization.  On the other hand, for spacetimes with a hypersurface that has a boundary, the above
construction shows that unlike the ADT charge, which is defined on
the boundary, the Taub charge has a boundary and a bulk piece. Nevertheless
the values of the charges must be equal to each other up to a sign, as
in (\ref{equalityofcharges}).

In the next section, we provide an explicit form of the $(\ensuremath{{\cal {G}}}_{\mu\nu})^{\left(2\right)}$
and the current $\bar{\xi}^{\nu}(\ensuremath{{\cal {G}}}_{\mu\nu})^{\left(2\right)}$ in terms of the perturbation $h$
which is another way to understand our more compact formulation. 

\section{Taub charges in the transverse-traceless gauge}

Consider a generic Einstein space $\bar{g}$ as the background with
\begin{equation}
\bar{R}_{\mu\nu}=\frac{2\Lambda}{n-2}\bar{g}_{\mu\nu},~~~~~~\bar{R}=\frac{2\Lambda n}{n-2}.
\end{equation}
Assuming we have  the first order field equations $(\ensuremath{{\cal {G}}}_{\mu\nu})^{\left(1\right)}=0$
which yields $\left(R\right)^{\left(1\right)}=0$ and 
\begin{equation}
\left(R_{\mu\nu}\right)^{\left(1\right)}=\frac{2\Lambda}{n-2}h_{\mu\nu}.
\end{equation}
The second order cosmological Einstein tensor
\begin{equation}
(\ensuremath{{\cal {G}}}_{\mu\nu}){}^{\left(2\right)}=(R_{\mu\nu}){}^{\left(2\right)}-\frac{1}{2}\bar{g}_{\mu\nu}(R){}^{\left(2\right)}-\frac{1}{2}h_{\mu\nu}(R){}^{\left(1\right)},
\end{equation}
  upon use of the first order equations
becomes
\begin{equation}
(\ensuremath{{\cal {G}}}_{\mu\nu}){}^{\left(2\right)}=(R_{\mu\nu}){}^{\left(2\right)}-\frac{1}{2}\bar{g}_{\mu\nu}(R){}^{\left(2\right)},
\end{equation}
where the second order Ricci tensor reads 
\begin{equation}
(R_{\mu\nu})^{(2)}=\bar{\nabla}_{\rho}(\Gamma_{\nu\mu}^{\rho})^{(2)}-\bar{\nabla}_{\nu}(\Gamma_{\rho\mu}^{\rho})^{(2)}+(\Gamma_{\mu\nu}^{\alpha})^{(1)}(\Gamma_{\sigma\alpha}^{\sigma})^{(1)}-(\Gamma_{\mu\sigma}^{\alpha})^{(1)}(\Gamma_{\nu\alpha}^{\sigma})^{(1)}.
\end{equation}
More explicitly, one has 
\begin{eqnarray}
 &  & (R_{\mu\nu})^{(2)}=-\frac{1}{2}\bar{\nabla}_{\sigma}\left(h^{\sigma\beta}\left(\bar{\nabla}_{\nu}h_{\mu\beta}+\bar{\nabla}_{\mu}h_{\nu\beta}-\bar{\nabla}_{\beta}h_{\mu\nu}\right)\right)+\frac{1}{4}\bar{\nabla}_{\nu}\bar{\nabla}_{\mu}(h_{\alpha\beta}h^{\alpha\beta})-\frac{1}{4}\bar{\nabla}_{\nu}h^{\alpha\beta}\bar{\nabla}_{\mu}h_{\alpha\beta}\nonumber \\
 &  &\hskip 1.8 cm +\frac{1}{4}\bar{\nabla}^{\sigma}h\left(\bar{\nabla}_{\nu}h_{\mu\sigma}+\bar{\nabla}_{\mu}h_{\nu\sigma}-\bar{\nabla}_{\sigma}h_{\mu\nu}\right)-\frac{1}{2}\bar{\nabla}^{\sigma}h_{\mu\alpha}\bar{\nabla}^{\alpha}h_{\nu\sigma}+\frac{1}{2}\bar{\nabla}^{\sigma}h_{\mu\alpha}\bar{\nabla}_{\sigma}h_{\nu}^{\alpha}.\label{secondorderricci}
\end{eqnarray}
From now on we will work in a specific gauge to simplify the computations.
The transverse-traceless (TT) gauge, $\bar{\nabla}_{\mu}h^{\mu\nu}=0$ and
$h=0$, is compatible with the field equations $(\ensuremath{{\cal {G}}}_{\mu\nu})^{\left(1\right)}=0$, which now read
\begin{equation}
\bar{\square}h_{\mu\nu}=2\bar{R}_{\alpha\mu\nu\beta}h^{\alpha\beta}.
\end{equation}
In the TT gauge one has
\begin{eqnarray}
 &  & (R_{\mu\nu})^{(2)}=-\frac{1}{2}\bar{\nabla}_{\sigma}\left(h^{\sigma\beta}\left(\bar{\nabla}_{\nu}h_{\mu\beta}+\bar{\nabla}_{\mu}h_{\nu\beta}-\bar{\nabla}_{\beta}h_{\mu\nu}\right)\right)+\frac{1}{4}\bar{\nabla}_{\nu}\bar{\nabla}_{\mu}(h_{\alpha\beta}h^{\alpha\beta})\nonumber \\
 &  & \hskip 2 cm  -\frac{1}{4}\bar{\nabla}_{\nu}h^{\alpha\beta}\bar{\nabla}_{\mu}h_{\alpha\beta}-\frac{1}{2}\bar{\nabla}^{\sigma}h_{\mu\alpha}\bar{\nabla}^{\alpha}h_{\nu\sigma}+\frac{1}{2}\bar{\nabla}^{\sigma}h_{\mu\alpha}\bar{\nabla}_{\sigma}h_{\nu}^{\alpha}.\label{secondorderricciinttgauge}
\end{eqnarray}
Straightforward manipulations yield 
\begin{eqnarray}
 &  & (R_{\mu\nu})^{(2)}=
  \frac{1}{4}h^{\alpha\beta}\bar{\nabla}_{\nu}\bar{\nabla}_{\mu}h_{\alpha\beta}+\frac{1}{2}\bar{\nabla}_{\sigma}h_{\mu\beta}\bar{\nabla}_{\nu}h^{\sigma\beta}+\frac{1}{2}\bar{\nabla}_{\sigma}h_{\nu\beta}\bar{\nabla}_{\mu}h^{\sigma\beta}+\frac{3\varLambda}{n-2}h_{\mu\beta}h_{\nu}^{\beta}\nonumber \\
 &  & +\frac{1}{4}h^{\lambda\sigma}(h_{\mu}^{\alpha}\bar{R}_{\sigma\alpha\nu\lambda}+h_{\nu}^{\alpha}\bar{R}_{\sigma\alpha\mu\lambda})  \\
&&+\frac{1}{4}\bar{\nabla}_{\sigma}\bar{\nabla}_{\lambda}(2h^{\sigma\lambda}h_{\mu\nu}-2\delta_{\nu}^{\lambda}h_{\mu\beta}h^{\sigma\beta}-2\delta_{\mu}^{\lambda}h^{\sigma\beta}h_{\nu\beta}+\frac{1}{2}\delta_{\nu}^{\sigma}\delta_{\mu}^{\lambda}h_{\alpha\beta}^{2}-h_{\mu}^{\lambda}h_{\nu}^{\sigma}-h_{\mu}^{\sigma}h_{\nu}^{\lambda}+\bar{g}^{\sigma\lambda}h_{\alpha\mu}h_{\nu}^{\alpha}).\nonumber \\
\label{secondorderricciasbulkandboundary} \nonumber
\end{eqnarray}
The second order perturbation of the scalar curvature
\begin{equation}
(R)^{(2)}=\bar{R}_{\mu\nu}h_{\alpha}^{\mu}h^{\alpha\nu}-(R_{\mu\nu})^{(1)}h^{\mu\nu}+\bar{g}^{\mu\nu}(R_{\mu\nu})^{(2)},
\end{equation}
reduces to 
\begin{equation}
(R)^{(2)}=\bar{\nabla}_{\sigma}\bar{\nabla}_{\lambda}\left(\frac{3}{8}\bar{g}^{\sigma\lambda}h_{\alpha\beta}^{2}-\frac{1}{2}h_{\rho}^{\lambda}h^{\rho\sigma}\right)+\frac{\varLambda}{n-2}h_{\alpha\beta}h^{\alpha\beta}.
\end{equation}
Combining the above results we can express the second order cosmological Einstein tensor as a divergence and a residual part as
\begin{equation}
(\ensuremath{{\cal {G}}}_{\mu\nu}){}^{\left(2\right)}=\bar{\nabla}_{\sigma}\bar{\nabla}_{\lambda}\mathcal{F}^{\sigma\lambda}\thinspace_{\mu\nu}+Y_{\mu\nu},
\end{equation}
where $\mathcal{F}^{\sigma\lambda}\thinspace_{\mu\nu}$ and $Y_{\mu\nu}$
are both symmetric in $\mu$ and $\nu$. Here the $\mathcal{F}$-tensor
reads
\begin{eqnarray}
 &  & \mathcal{F}^{\sigma\lambda}\thinspace_{\mu\nu}=\frac{1}{2}h^{\sigma\lambda}h_{\mu\nu}-\delta_{(\mu}^{\lambda}h_{\nu)\beta}h^{\sigma\beta}+\frac{1}{8}\delta_{\nu}^{\sigma}\delta_{\mu}^{\lambda}h_{\alpha\beta}^{2}-\frac{1}{2}h_{(\mu}^{\lambda}h_{\nu)}^{\sigma}\nonumber \\
 &  &  \hskip 1.4 cm +\frac{1}{4}\bar{g}^{\sigma\lambda}h_{\alpha\mu}h_{\nu}^{\alpha}-\frac{3}{16}\bar{g}_{\mu\nu}\bar{g}^{\sigma\lambda}h_{\alpha\beta}h^{\alpha\beta}+\frac{1}{4}\bar{g}_{\mu\nu}h_{\rho}^{\lambda}h^{\rho\sigma},\label{boundarytermofsecondorderequation}
\end{eqnarray}
and the $Y$-tensor reads
\begin{eqnarray}
 &  & Y_{\mu\nu}=\frac{1}{2}h^{\alpha\beta}\bar{\nabla}_{(\mu}\bar{\nabla}_{\nu)}h_{\alpha\beta}+\bar{\nabla}_{\sigma}h_{\beta (\mu}\bar{\nabla}_{\nu)}h^{\sigma\beta}+\frac{3\varLambda}{n-2}h_{\mu\beta}h_{\nu}^{\beta}\nonumber \\
 &  & \hskip 1 cm +\frac{1}{2}h^{\lambda\sigma}h_{(\mu}^{\alpha}\bar{R}_{\nu) \lambda\sigma\alpha}-\frac{\varLambda}{2(n-2)}\bar{g}_{\mu\nu}h_{\alpha\beta}^2.\label{bulkofsecondorderequation}
\end{eqnarray}
So $(\ensuremath{{\cal {G}}}_{\mu\nu}){}^{\left(2\right)}$ has a
divergence part and a part which is not of the divergence type. One
can further try to manipulate the $Y_{\mu\nu}$ to obtain some
divergence terms, but one always ends up with terms which cannot be written
as a divergence of any tensor as expected. Let $\bar{\xi}$ be a background Killing vector field. Contraction
with the second order perturbation of the cosmological Einstein tensor
yields
\begin{equation}
\bar{\xi}^{\nu}(\ensuremath{{\cal {G}}}_{\mu\nu}){}^{\left(2\right)}=\bar{\nabla}_{\sigma}\left(\bar{\xi}^{\nu}\bar{\nabla}_{\lambda}\mathcal{F}^{\sigma\lambda}\thinspace_{\mu\nu}-\mathcal{F}^{\lambda\sigma}\thinspace_{\mu\nu}\bar{\nabla}_{\lambda}\bar{\xi}^{\nu}\right)+\mathcal{F}^{\sigma\lambda}\thinspace_{\mu\nu}\bar{\nabla}_{\lambda}\bar{\nabla}_{\sigma}\bar{\xi}^{\nu}+Y_{\mu\nu}\bar{\xi}^{\nu}.
\end{equation}
In background {\it Einstein spaces}, the last two terms can be written
as 
\begin{eqnarray}
 &  & \mathcal{F}^{\sigma\lambda}\thinspace_{\mu\nu}\bar{\nabla}_{\lambda}\bar{\nabla}_{\sigma}\bar{\xi}^{\nu}+Y_{\mu\nu}\bar{\xi}^{\nu}=\frac{1}{4}\bar{\xi}^{\nu}h^{\alpha\beta}\bar{\nabla}_{\nu}\bar{\nabla}_{\mu}h_{\alpha\beta}+\frac{1}{2}\bar{\xi}^{\nu}\bar{\nabla}_{\sigma}h_{\mu\beta}\bar{\nabla}_{\nu}h^{\sigma\beta}+\frac{1}{2}\bar{\xi}^{\nu}\bar{\nabla}_{\sigma}h_{\nu\beta}\bar{\nabla}_{\mu}h^{\sigma\beta}\nonumber \\
 &  & +\frac{3\varLambda}{2(n-2)}\bar{\xi}^{\nu}h_{\mu\beta}h_{\nu}^{\beta}-\frac{\varLambda}{8(n-2)}\bar{\xi}_{\mu}h_{\alpha\beta}^{2}+\frac{1}{4}\bar{\xi}^{\nu}h^{\lambda\sigma}h_{\nu}^{\alpha}\bar{R}_{\sigma\alpha\mu\lambda}\nonumber \\
 &  & +\frac{1}{4}\bar{\xi}^{\rho}h^{\beta\sigma}h_{\beta}^{\lambda}\bar{R}_{\rho\lambda\sigma\mu}+\bar{\xi}^{\nu}h^{\lambda\sigma}h_{\mu}^{\alpha}\bar{R}_{\sigma\alpha\nu\lambda}.\label{lasttwotermsinbackgroundeinstein}
\end{eqnarray}
The important point is that unlike the case of the first order cosmological Einstein tensor as discussed after (\ref{onemli}), at the second order the residual parts as given in the last expression do not vanish upon use of the background and first order field equations. To see this more explicitly, let us look at the (anti) de Sitter and flat backgrounds.  In (anti)-de Sitter backgrounds one has 
\begin{eqnarray}
 Y_{\mu\nu}=\frac{1}{2}h^{\alpha\beta}\bar{\nabla}_{(\mu}\bar{\nabla}_{\nu)}h_{\alpha\beta}+\bar{\nabla}_{\sigma}h_{\beta (\mu}\bar{\nabla}_{\nu)}h^{\sigma\beta}
 +\frac{\varLambda(3n-2)}{(n-1)\left(n-2\right)}h_{\mu\beta}h_{\nu}^{\beta}-\frac{\varLambda}{2(n-2)}\bar{g}_{\mu\nu}h_{\alpha\beta}^{2},\label{bulkinads}
\end{eqnarray}
and the residual part is
\begin{eqnarray}
 &  & \mathcal{F}^{\sigma\lambda}\thinspace_{\mu\nu}\bar{\nabla}_{\lambda}\bar{\nabla}_{\sigma}\bar{\xi}^{\nu}+Y_{\mu\nu}\bar{\xi}^{\nu}=\frac{1}{4}\bar{\xi}^{\nu}h^{\alpha\beta}\bar{\nabla}_{\nu}\bar{\nabla}_{\mu}h_{\alpha\beta}+\frac{1}{2}\bar{\xi}^{\nu}\bar{\nabla}_{\sigma}h_{\mu\beta}\bar{\nabla}_{\nu}h^{\sigma\beta}+\frac{1}{2}\bar{\xi}^{\nu}\bar{\nabla}_{\sigma}h_{\nu\beta}\bar{\nabla}_{\mu}h^{\sigma\beta}\nonumber \\
 &  & +\frac{2\varLambda}{(n-1)(n-2)}\left(\frac{3}{4}(n+1)\bar{\xi}^{\nu}h_{\mu\beta}h_{\nu}^{\beta}-\frac{1}{16}(n+3)\bar{\xi}_{\mu}h_{\alpha\beta}^{2}\right).\label{lasttwotermsinAdS}
\end{eqnarray}
One realizes that no amount of manipulations can turn these terms
into a pure divergence. This is consistent with our compact expression
of the previous section. For example for flat spaces, considering $\Lambda=0$, with $\bar{\nabla}_{\mu}\rightarrow\partial_{\mu}$,
one can easily see that one has the non divergence part reads
\begin{equation}
\mathcal{F}^{\sigma\lambda}\thinspace_{\mu\nu}\partial_{\lambda}\partial_{\sigma}\bar{\xi}^{\nu}+Y_{\mu\nu}\bar{\xi}^{\nu}=\frac{1}{4}\bar{\xi}^{\nu}h^{\alpha\beta}\partial_{\nu}\partial_{\mu}h_{\alpha\beta}+\frac{1}{2}\bar{\xi}^{\nu}\partial_{\sigma}h_{\mu\beta}\partial_{\nu}h^{\sigma\beta}+\frac{1}{2}\bar{\xi}^{\nu}\partial_{\sigma}h_{\nu\beta}\partial_{\mu}h^{\sigma\beta},
\end{equation}
which cannot be written as a pure divergence. 

\section{Gauge invariance issue}

The first order linearized cosmological Einstein tensor is gauge invariant for Einstein metrics under small gauge transformations, but the second order cosmological Einstein tensor is not. Therefore, it pays to lay out some of the details of these and the gauge transformation properties of the tensors and currents we have constructed.  
Under a gauge transformation generated by a vector field $X$, the
first order metric perturbation changes as
\begin{equation}
\delta_{X}h_{\mu\nu}=\bar{\nabla}_{\nu}X_{\mu}+\bar{\nabla}_{\mu}X_{\nu}.
\end{equation}
As noted above, it is easy to see that $({\cal {G}}_{\mu\nu})^{\left(1\right)}$ is
gauge invariant once the background space is an Einstein space. But
$({\cal {G}}_{\mu\nu})^{\left(2\right)}$ is not gauge invariant,
in fact a pure divergence part is generated. Let us show this in a
systematic way following \cite{Fischer-Marsden-Moncrief}. Let $\lambda~\in~\mathbb{R}$ and $\varphi$
be a one parameter family of diffeomorphisms acting on the spacetime manifold
$\varphi:\mathbb{R}\times\mathscr{M}\rightarrow\text{\ensuremath{\mathscr{M}}}$,
then diffeomorphism invariance of a tensor field $T$ means
\begin{equation}
T(\varphi^{*}g)=\varphi^{*}T(g),\label{diffeomorphisminvariance}
\end{equation}
where $\varphi^{*}$ is the pullback map.  Let us denote the diffeomorphism by $\varphi_{\lambda}$ and assuming $\varphi_{0}$
to be the identity map. Differentiating (\ref{diffeomorphisminvariance}) with respect to
$\lambda$ yields
\begin{equation}
\frac{d}{d\lambda}T(\varphi_\lambda^{*}g)=\frac{d}{d\lambda}\varphi_\lambda^{*}T(g).
\end{equation}
Using the chain rule one has 
\begin{equation}
DT(\varphi_{\lambda}^{*}g)\cdot\frac{d}{d\lambda}\varphi_{\lambda}^{*}g=\varphi_{\lambda}^{*}\left(\mathscr{L}_{X}T(g)\right),\label{firstderivative}
\end{equation}
where $D$ denotes the Fr\'echet derivative and $\mathscr{L}_{X}$ denotes the Lie derivative along the vector field $X$. In local coordinates for a rank $(0,2)$ tensor field-which is relevant for field equation-the last expression
yields
\begin{equation}
\delta_{X}(T_{\mu\nu})^{\left(1\right)}\cdot h=\mathscr{L}_{X}\bar{T}_{\mu\nu}.
\end{equation}
Specifically for the cosmological Einstein tensor $T_{\mu\nu}={\cal {G}}_{\mu\nu}$, we have 
\begin{equation}
\delta_{X}({\cal {G}}_{\mu\nu})^{\left(1\right)}\cdot h=\mathscr{L}_{X}\bar{{\cal {G}}}_{\mu\nu}=0,
\end{equation}
which is a statement of the gauge invariance of the first order linearized cosmological Einstein tensor.  For the second order tensors, we can take another derivative of (\ref{firstderivative}) to get
\begin{equation}
D^{2}T(g)\cdot\left(h,\text{\ensuremath{\mathscr{L}}}_{X}g\right)+DT(g)\cdot\text{\ensuremath{\mathscr{L}}}_{X}h=\text{\ensuremath{\mathscr{L}}}_{X}(DT(g)\cdot h),
\end{equation}
which yields in local coordinates 
\begin{equation}
\delta_{X}(T_{\mu\nu})^{\left(2\right)}\cdot[h,h]+(T_{\mu\nu})^{\left(1\right)}\cdot\mathscr{L}_{X}h=\text{\ensuremath{\mathscr{L}}}_{X}(T_{\mu\nu})^{\left(1\right)}\cdot h.
\end{equation}
 When $T_{\mu\nu}={\cal {G}}_{\mu\nu}$, we obtain
\begin{equation}
\delta_{X}({\cal {G}}_{\mu\nu})^{\left(2\right)}\cdot[h,h]+({\cal {G}}_{\mu\nu})^{\left(1\right)}\cdot\mathscr{L}_{X}h=\text{\ensuremath{\mathscr{L}}}_{X}({\cal {G}}_{\mu\nu})^{\left(1\right)}\cdot h.
\end{equation}
The right-hand side is zero for linearized solutions; and one obtains 
\begin{equation}
\delta_{X}({\cal {G}}_{\mu\nu})^{\left(2\right)}\cdot[h,h]=-({\cal {G}}_{\mu\nu})^{\left(1\right)}\cdot\mathscr{L}_{X}h.
\end{equation}
The right-hand side of this expression is not zero but it can be written as a pure divergence term proving our earlier claim. We give a more direct, albeit highly cumbersome derivation of this expression in  Appendix B 
using the explicit form of $({\cal {G}}_{\mu\nu})^{\left(2\right)}$. 

Let us now study the gauge transformation of (\ref{secondordermainequation})
and see explicitly that the right-hand side is a pure boundary. The first order linearized $(\text{\ensuremath{{\cal {P}}}}$ tensor reads
\begin{equation}
(\text{\ensuremath{{\cal {P}}}}^{\nu\mu}\thinspace_{\beta\sigma})^{\left(1\right)}=(R^{\nu\mu}\thinspace_{\beta\sigma})^{\left(1\right)},
\end{equation}
which is gauge invariant under the small coordinate transformations for (anti) de Sitter backgrounds as it can be seen from (\ref{firstderivative}).
Defining $c=\frac{(n-1)(n-2)}{4\Lambda(n-3)}$, we have
\begin{equation}
\frac{1}{c}\bar{\xi}^{\nu}\delta_{X}(\ensuremath{{\cal {G}}}_{\nu}^{\mu})^{\left(2\right)}=\bar{\nabla}_{\nu}\left(\bar{\text{\ensuremath{{\cal {F}}}}}^{\beta\sigma}\delta_{X}(\text{\ensuremath{{\cal {P}}}}^{\nu\mu}\thinspace_{\beta\sigma})^{\left(2\right)}+\bar{\nabla}_{\lambda}X^{\lambda}\bar{\text{\ensuremath{{\cal {F}}}}}^{\beta\sigma}\text{\ensuremath{{\cal {P}}}}^{\nu\mu}\thinspace_{\beta\sigma})^{\left(1\right)}\right)-2\delta_{X}(\Gamma_{\nu\rho}^{\beta})^{\left(1\right)}\bar{\text{\ensuremath{{\cal {F}}}}}^{\rho\sigma}(\text{\ensuremath{{\cal {P}}}}^{\nu\mu}\thinspace_{\beta\sigma})^{\left(1\right)}. \label{ellidok}
\end{equation}
Since the first two terms are already boundary terms, let us consider
the last part:
\begin{equation}
\delta_{X}(\Gamma_{\nu\rho}^{\beta})^{\left(1\right)}\bar{\text{\ensuremath{{\cal {F}}}}}^{\rho\sigma}(\text{\ensuremath{{\cal {P}}}}^{\nu\mu}\thinspace_{\beta\sigma})^{\left(1\right)}=\left(\bar{\nabla}_{\nu}\bar{\nabla}_{\rho}X^{\beta}+\bar{R}^{\beta}\thinspace_{\rho\lambda\nu}X^{\lambda}\right)\bar{\text{\ensuremath{{\cal {F}}}}}^{\rho\sigma}(\text{\ensuremath{{\cal {P}}}}^{\nu\mu}\thinspace_{\beta\sigma})^{\left(1\right)},
\end{equation}
where we used (\ref{eq:gaugetranschristoffel}) of Appendix B. One
can rewrite this as
\begin{eqnarray}
 \delta_{X}(\Gamma_{\nu\rho}^{\beta})^{\left(1\right)}\bar{\text{\ensuremath{{\cal {F}}}}}^{\rho\sigma}(\text{\ensuremath{{\cal {P}}}}^{\nu\mu}\thinspace_{\beta\sigma})^{\left(1\right)} &  &=\bar{\nabla}_{\nu}\left(\bar{\text{\ensuremath{{\cal {F}}}}}^{\rho\sigma}(\text{\ensuremath{{\cal {P}}}}^{\nu\mu}\thinspace_{\beta\sigma})^{\left(1\right)}\bar{\nabla}_{\rho}X^{\beta}\right)-\bar{\nabla}_{\nu}\bar{\text{\ensuremath{{\cal {F}}}}}^{\rho\sigma}(\text{\ensuremath{{\cal {P}}}}^{\nu\mu}\thinspace_{\beta\sigma})^{\left(1\right)}\bar{\nabla}_{\rho}X^{\beta}\nonumber \\
 &  & +\frac{2\varLambda}{(n-1)(n-2)}X^{\beta}\bar{\text{\ensuremath{{\cal {F}}}}}_{\nu}\thinspace^{\sigma}(\text{\ensuremath{{\cal {P}}}}^{\nu\mu}\thinspace_{\beta\sigma})^{\left(1\right)}.\label{gaugeinvarianceofnondivergenceterm}
\end{eqnarray}
By using $\bar{\nabla}_{\nu}\bar{\text{\ensuremath{{\cal {F}}}}}^{\rho\sigma}=\bar{R}_{\gamma\nu}\thinspace^{\rho\sigma}\bar{\xi}^{\gamma}$,
one has
\begin{equation}
\delta_{X}(\Gamma_{\nu\rho}^{\beta})^{\left(1\right)}\bar{\text{\ensuremath{{\cal {F}}}}}^{\rho\sigma}(\text{\ensuremath{{\cal {P}}}}^{\nu\mu}\thinspace_{\beta\sigma})^{\left(1\right)}=\bar{\nabla}_{\nu}\left((\text{\ensuremath{{\cal {P}}}}^{\nu\mu}\thinspace_{\beta\sigma})^{\left(1\right)}\left(\bar{\text{\ensuremath{{\cal {F}}}}}^{\rho\sigma}\bar{\nabla}_{\rho}X^{\beta}+\frac{2\varLambda}{(n-1)(n-2)}\bar{\xi}^{\sigma}X^{\beta}\right)\right). \label{altmisiki}
\end{equation}
Therefore the Taub current is not gauge invariant as expected, under gauge transformations a boundary part which is composed of the first part of (\ref{ellidok}) and (\ref{altmisiki}) whose divergence vanishes, is generated. 

\section{{\normalsize{}{}{}{}Conclusions}}

In a nonlinear theory, validity of perturbation theory about an exact
solution is a subtle issue. In General Relativity, if the background
metric $\bar{g}$, about which perturbation theory is performed, has
Killing symmetries, there are constraints to the first order perturbation
theory coming from the second order perturbation theory. We have explicitly
studied the constraints and have shown that the Taub charge, which
is an integral constraint on the first order perturbation, does not
vanish automatically. We have identified the bulk and boundary terms
in the conserved current $\sqrt{-\bar{g}}\bar{\xi}_{\mu}(\ensuremath{{\cal {G}}}^{\mu\nu})^{(2)}\cdot[h,h]$.
This issue is quite important when one looks for the perturbative
solutions in spacetimes with closed hypersurfaces and it is also
relevant for semi-classical quantization of gravity in such backgrounds. 

From another vantage point, one can understand these results as follows:
the solution space of Einstein equations generically form a manifold
except at solutions $\bar{g}$ that have Killing fields. Around
such a metric $\bar{g}$, the linearized field equations which yield
the tangent space of the solution space give a larger dimensional space.
Therefore the linearized solutions yield some nonintegrable deformations.
One pays this at the second order where there is a constraint on the first
order solutions. 
\section{Appendix A: Second order perturbation theory }

Let us summarize some results about the second order perturbation
theory (see also \cite{sismanBI}). Assuming $\bar{g}_{\mu\nu}$ to be a generic background metric, by definition one has 
\begin{equation}
g_{\mu\nu}:=\bar{g}_{\mu\nu}+\lambda h_{\mu\nu},
\end{equation}
with an inverse  
\begin{equation}
g^{\mu\nu}=\bar{g}^{\mu\nu}-\lambda h^{\mu\nu}+\lambda^{2}h_{\alpha}^{\mu}h^{\alpha\nu}+{\cal{O}}(\lambda^{3}).
\end{equation}
Let $T$ be a generic tensor depending on the metric, then it can be expanded as 
\begin{equation}
T=\bar{T}+\lambda{T}^{(1)}+\lambda^{2}{T}^{(2)}+{\cal{O}}(\lambda^{3}).
\end{equation}
The Christoffel connection reads 
\begin{equation}
\Gamma_{\mu\nu}^{\gamma}=\bar{\Gamma}_{\mu\nu}^{\gamma}+\lambda(\Gamma_{\mu\nu}^{\gamma})^{(1)}+\lambda^{2}(\Gamma_{\mu\nu}^{\gamma})^{(2)},
\end{equation}
where the first order term is 
\begin{equation}
(\Gamma_{\mu\nu}^{\gamma})^{(1)}=\frac{1}{2}\big(\bar{\nabla}_{\mu}h_{\nu}^{\gamma}+\bar{\nabla}_{\nu}h_{\mu}^{\gamma}-\bar{\nabla}^{\gamma}h_{\mu\nu}\big),
\end{equation}
and the second order expansion is
\begin{equation}
(\Gamma_{\mu\nu}^{\gamma})^{(2)}=-h_{\delta}^{\gamma}(\Gamma_{\mu\nu}^{\delta})^{(1)}.
\end{equation}
Since it is a background tensor, we can raise and lower its indices
with $\bar{g}_{\mu\nu}$. Our definition is  
\begin{equation}
(\Gamma_{\mu\nu\delta})^{(1)} :=\bar{g}_{\gamma\delta}(\Gamma_{\mu\nu}^{\gamma})^{(1)}.
\end{equation}
The first order linearized Riemann tensor is 
\begin{equation}
(R^{\rho}\thinspace_{\mu\sigma\nu})^{(1)}=\bar{\nabla}_{\sigma}(\Gamma_{\nu\mu}^{\rho})^{(1)}-\bar{\nabla}_{\nu}(\Gamma_{\sigma\mu}^{\rho})^{(1)},
\end{equation}
and the second order linearized Riemann tensor is 
\begin{equation}
(R^{\rho}\thinspace_{\mu\sigma\nu})^{(2)}=\bar{\nabla}_{\sigma}(\Gamma_{\nu\mu}^{\rho})^{(2)}-\bar{\nabla}_{\nu}(\Gamma_{\sigma\mu}^{\rho})^{(2)}-(\Gamma_{\mu\nu}^{\alpha})^{(1)}(\Gamma_{\sigma\alpha}^{\rho})^{(1)}+(\Gamma_{\mu\sigma}^{\alpha})^{(1)}(\Gamma_{\nu\alpha}^{\rho})^{(1)}.
\end{equation}
The first order linearized Ricci tensor is 
\begin{equation}
(R_{\mu\nu})^{(1)}=\bar{\nabla}_{\sigma}(\Gamma_{\mu\nu}^{\sigma})^{(1)}-\bar{\nabla}_{\nu}(\Gamma_{\sigma\mu}^{\sigma})^{(1)},
\end{equation}
and the second order linearized Ricci tensor is 
\begin{equation}
(R_{\mu\nu})^{(2)}=\bar{\nabla}_{\sigma}(\Gamma_{\nu\mu}^{\sigma})^{(2)}-\bar{\nabla}_{\nu}(\Gamma_{\sigma\mu}^{\sigma})^{(2)}-(\Gamma_{\mu\nu}^{\alpha})^{(1)}(\Gamma_{\sigma\alpha}^{\sigma})^{(1)}+(\Gamma_{\mu\sigma}^{\alpha})^{(1)}(\Gamma_{\nu\alpha}^{\sigma})^{(1)}.\label{secondorderricciintermsofconnection}
\end{equation}
The linearized scalar curvature is 
\begin{equation}
(R)^{(1)}=\bar{\nabla}_{\alpha}\bar{\nabla}_{\beta}h^{\alpha\beta}-\bar{\square}h-\bar{R}_{\mu\nu}h^{\mu\nu},
\end{equation}
and the second order linearized scalar curvature is 
\begin{equation}
(R)^{(2)}=\bar{R}_{\mu\nu}h_{\alpha}^{\mu}h^{\alpha\nu}-(R_{\mu\nu})^{(1)}h^{\mu\nu}+\bar{g}^{\mu\nu}(R_{\mu\nu})^{(2)}.\label{secondordercurvaturescalarintermsofconnection}
\end{equation}
The cosmological Einstein tensor 
\begin{equation}
\ensuremath{{\cal {G}}}_{\mu\nu}=R_{\mu\nu}-\frac{1}{2}g_{\mu\nu}R+\Lambda g_{\mu\nu},
\end{equation}
at second order reads
\begin{equation}
(\ensuremath{{\cal {G}}}_{\mu\nu}){}^{\left(2\right)}=(R_{\mu\nu}){}^{\left(2\right)}-\frac{1}{2}\bar{g}_{\mu\nu}(R){}^{\left(2\right)}-\frac{1}{2}h_{\mu\nu}(R){}^{\left(1\right)}.\label{eq:secondordercosmologicaleinstein-1}
\end{equation}
We have already given the first order form of the cosmological Einstein tensor in section II.

\section {APPENDIX B: Gauge Transformations }

Lie and covariant derivatives do not commute; but, sometimes
we need to change the order of the these two differentiations. First we provide some identities which can be easily proven from the definitions. Under a gauge transformation, $\delta_{X}h_{\mu \nu}=\bar{\nabla}_{\mu}X_{\nu}+\bar{\nabla}_{\nu}X_{\mu}$, one has
\begin{equation}
\delta_{X}(\Gamma_{\mu \nu}^{\sigma}){}^{\left(1\right)}=\frac{1}{2}\bigg(\bar{\nabla}_{\mu}\delta_{X}h_{\nu}^{\sigma}+\bar{\nabla}_{\nu}\delta_{X}h_{\mu}^{\sigma}-\bar{\nabla}^{\sigma}\delta_{X}h_{\mu \nu}\bigg),
\end{equation}
which yields
\begin{equation}
\delta_{X}(\Gamma_{\mu \nu}^{\sigma}){}^{\left(1\right)}=\bar{\nabla}_{\mu}\bar{\nabla}_{\nu}X^{\sigma}+\bar{R}^{\sigma}\thinspace_{\nu \rho \mu }X^{\rho}.\label{eq:gaugetranschristoffel}
\end{equation}
We used this form in the text.  

For a generic rank $(m,n)$ tensor field, one can prove the following expression:
\begin{eqnarray}
 &  & \bar{\nabla}_{\sigma}\text{\ensuremath{\mathscr{L}}}_{X}T^{\nu_{1}\nu_{2}...\nu_{m}}\thinspace_{\mu_{1}\mu_{2}...\mu_{n}}=\text{\ensuremath{\mathscr{L}}}_{X}\bar{\nabla}_{\sigma}T^{\nu_{1}\nu_{2}...\nu_{m}}\thinspace_{\mu_{1}\mu_{2}...\mu_{n}}\label{eq:identitygeneral}\\
 &  & +\delta_{X}(\Gamma_{\sigma \mu_{1}}^{\rho})^{\left(1\right)}T^{\nu_{1}\nu_{2}...\nu_{m}}\thinspace_{\rho \mu_{2}...\mu_{n}}+\delta_{X}(\Gamma_{\sigma \mu_{2}}^{\rho })^{\left(1\right)}T^{\nu_{1}\nu_{2}...\nu_{m}}\thinspace_{\mu_{1}\rho ...\mu_{n}}+...+\delta_{X}(\Gamma_{ \sigma \mu_{n}}^{\rho })^{\left(1\right)}T^{\nu_{1}\nu_{2}...\nu_{m}}\thinspace_{\mu_{1}\mu_{2}...\rho}\nonumber \\
 &  & -\delta_{X}(\Gamma_{ \sigma \rho }^{\nu_{1}})^{\left(1\right)}T^{\rho \nu_{2}...\nu_{m}}\thinspace_{\mu_{1}\mu_{2}...\mu_{n}}-\delta_{X}(\Gamma_{\sigma \rho }^{\nu_{2}})^{\left(1\right)}T^{\nu_{1}\rho ...\nu_{m}}\thinspace_{\mu_{1}\mu_{2}...\mu_{n}}-...-\delta_{X}(\Gamma_{ \sigma \rho}^{\nu_{m}})^{\left(1\right)}T^{\nu_{1}\nu_{2}...\rho}\thinspace_{\mu_{1}\mu_{2}...\mu_{n}}.\nonumber 
\end{eqnarray}
The second order Ricci tensor (\ref{secondorderricciintermsofconnection}) transforms as
\begin{eqnarray}
 &  & \delta_{X}(R{}_{\mu\nu})^{(2)}=-\bar{\nabla}_{\rho}\left(\delta_{X}h_{\beta}^{\rho}(\Gamma_{\nu\mu}^{\beta})^{(1)}+h_{\beta}^{\rho}\delta_{X}(\Gamma_{\nu\mu}^{\beta})^{(1)}\right)+\bar{\nabla}_{\nu}\left(\delta_{X}h_{\beta}^{\rho}(\Gamma_{\rho\mu}^{\beta})^{(1)}+h_{\beta}^{\rho}\delta_{X}(\Gamma_{\rho\mu}^{\beta})^{(1)}\right)\nonumber \\
 &  &  \hskip 2.5 cm +\delta_{X}\left((\Gamma_{\mu\nu}^{\alpha})^{(1)}(\Gamma_{\sigma\alpha}^{\sigma})^{(1)}-(\Gamma_{\mu\sigma}^{\alpha})^{(1)}(\Gamma_{\nu\alpha}^{\sigma})^{(1)}\right).\label{gaugetransformationofsecondorderrricci2}
\end{eqnarray}
Using $\delta_{X}h^{\rho\beta}=-\text{\ensuremath{\mathscr{L}}}_{X}\bar{g}^{\rho\beta}$, one has 
\begin{equation}
\delta_{X}h_{\beta}^{\rho}(\Gamma_{\nu\mu}^{\beta})^{(1)}=-\text{\ensuremath{\mathscr{L}}}_{X}\bar{g}^{\rho\beta}(\Gamma_{\nu\mu\beta})^{(1)}.
\end{equation}
Then
\begin{eqnarray}
 &  & \delta_{X}(R{}_{\mu\nu})^{(2)}=\bar{\nabla}_{\rho}\left(\text{\ensuremath{\mathscr{L}}}_{X}\bar{g}^{\rho\beta}(\Gamma_{\nu\mu\beta})^{(1)}\right)-\bar{\nabla}_{\nu}\left(\text{\ensuremath{\mathscr{L}}}_{X}\bar{g}^{\rho\beta}(\Gamma_{\rho\mu\beta})^{(1)}\right)+\bar{\nabla}_{\nu}\left(h_{\beta}^{\rho}\delta_{X}(\Gamma_{\rho\mu}^{\beta})^{(1)}\right)\nonumber \\
 &  & \hskip 2 cm -\bar{\nabla}_{\rho}\left(h_{\beta}^{\rho}\delta_{X}(\Gamma_{\nu\mu}^{\beta})^{(1)}\right)+\delta_{X}\left((\Gamma_{\mu\nu}^{\alpha})^{(1)}(\Gamma_{\sigma\alpha}^{\sigma})^{(1)}-(\Gamma_{\mu\sigma}^{\alpha})^{(1)}(\Gamma_{\nu\alpha}^{\sigma})^{(1)}\right).\label{gaugetransformationofricci2}
\end{eqnarray}
 After using the identity (\ref{eq:identitygeneral}),  one gets
\begin{eqnarray}
 &  & \delta_{X}(R{}_{\mu\nu})^{(2)}=\text{\ensuremath{\mathscr{L}}}_{X}(R_{\mu\nu})^{(1)}-\frac{1}{2}\bar{g}^{\rho\beta}\bar{\nabla}_{\rho}\text{\ensuremath{\mathscr{L}}}_{X}\left(\bar{\nabla}_{\nu}h_{\mu\beta}+\bar{\nabla}_{\mu}h_{\nu\beta}-\bar{\nabla}_{\beta}h_{\mu\nu}\right)\nonumber \\
 &  & \hskip 1.4 cm +\frac{1}{2}\bar{g}^{\rho\beta}\bar{\nabla}_{\nu}\text{\ensuremath{\mathscr{L}}}_{X}\bar{\nabla}_{\mu}h_{\rho\beta}+\bar{\nabla}_{\nu}\left(h_{\beta}^{\rho}\delta_{X}(\Gamma_{\rho\mu}\thinspace^{\beta})^{(1)}\right)-\bar{\nabla}_{\rho}\left(h_{\beta}^{\rho}\delta_{X}(\Gamma_{\nu\mu}\thinspace^{\beta})^{(1)}\right),\label{gaugetransformationofricci2_3}
\end{eqnarray}
which simplifies to 
\begin{eqnarray}
\delta_{X}(R{}_{\mu\nu})^{(2)}=&&\text{\ensuremath{\mathscr{L}}}_{X}(R_{\mu\nu})^{(1)} \\
&&-\frac{\bar{g}^{\rho\beta}}{2}\left(\bar{\nabla}_{\rho}\bar{\nabla}_{\nu}\text{\ensuremath{\mathscr{L}}}_{X}h_{\mu\beta}+\bar{\nabla}_{\rho}\bar{\nabla}_{\mu}\text{\ensuremath{\mathscr{L}}}_{X}h_{\nu\beta}-\bar{\nabla}_{\rho}\bar{\nabla}_{\beta}\text{\ensuremath{\mathscr{L}}}_{X}h_{\mu\nu}-\bar{\nabla}_{\nu}\bar{\nabla}_{\mu}\text{\ensuremath{\mathscr{L}}}_{X}h_{\rho\beta}\right), \nonumber 
\end{eqnarray}
where the last four terms yield the Ricci tensor evaluated at the Lie derivative
of the linear metric perturbation. Finally we can write
\begin{equation}
\delta_{X}(R{}_{\mu\nu})^{(2)}=\text{\ensuremath{\mathscr{L}}}_{X}(R_{\mu\nu})^{(1)}-\left(R_{\mu\nu}\right)^{(1)}\cdot\text{\ensuremath{\mathscr{L}}}_{X}h.
\end{equation}
For the gauge transformation of the second order linearized  scalar curvature
we need to compute 
\begin{equation}
\delta_{X}(R)^{(2)}=\bar{R}_{\mu\nu}\delta_{X}\left(h_{\alpha}^{\mu}h^{\alpha\nu}\right)-\delta_{X}(R_{\mu\nu})^{(1)}h^{\mu\nu}-(R_{\mu\nu})^{(1)}\delta_{X}h^{\mu\nu}+\bar{g}^{\mu\nu}\delta_{X}(R_{\mu\nu})^{(2)}.
\end{equation}
After a straightforward calculation, the result turns out to be
\begin{equation}
\delta_{X}(R)^{(2)}=\text{\ensuremath{\mathscr{L}}}_{X}(R)^{(1)}-\left(R\right)^{(1)}\cdot\text{\ensuremath{\mathscr{L}}}_{X}h.
\end{equation}
We can collect these pieces to write the gauge transformation of the second order cosmological Einstein tensor in a generic background as  
\begin{equation}
\delta_{X}({\cal {G}}_{\mu\nu})^{(2)}=\delta_{X}(R_{\mu\nu})^{(2)}-\frac{1}{2}\bar{g}_{\mu\nu}\delta_{X}(R)^{(2)}-\frac{1}{2}(R)^{(1)}\delta_{X}h_{\mu\nu}-\frac{1}{2}h_{\mu\nu}\delta_{X}(R)^{(1)}.
\end{equation}
Using the above results, the last expression becomes
\begin{equation}
\delta_{X}(\ensuremath{{\cal {G}}}_{\mu\nu})^{(2)}+(\ensuremath{{\cal {G}}}_{\mu\nu})^{(1)}\cdot\text{\ensuremath{\mathscr{L}}}_{X}h=\text{\ensuremath{\mathscr{L}}}_{X}(\ensuremath{{\cal {G}}}_{\mu\nu})^{(1)}.
\end{equation}
This result has been general and we have not used any field equations or their linearizations. 
When $h$ is solution to the first order linearized cosmological Einstein
tensor, the right hand side of the last expression vanishes, and we
have 
\begin{equation}
\delta_{X}(\ensuremath{{\cal {G}}}_{\mu\nu})^{(2)}=-(\ensuremath{{\cal {G}}}_{\mu\nu})^{(1)}\cdot\text{\ensuremath{\mathscr{L}}}_{X}h,
\end{equation}
which shows the gauge non-invariance of the second order cosmological Einstein tensor. 
Now let us consider the contraction of the result with a background
Killing vector field
\begin{equation}
\bar{\xi}^{\nu}\delta_{X}(\ensuremath{{\cal {G}}}_{\mu\nu})^{(2)}=-\bar{\xi}^{\nu}(\ensuremath{{\cal {G}}}_{\mu\nu})^{(1)}\cdot\text{\ensuremath{\mathscr{L}}}_{X}h,
\end{equation}
since $\bar{\xi}^{\nu}(\ensuremath{{\cal {G}}}_{\mu\nu})^{(1)}$ can
be expressed as a boundary term, $\bar{\xi}^{\nu}(\ensuremath{{\cal {G}}}_{\mu\nu})^{(1)}\cdot\text{\ensuremath{\mathscr{L}}}_{X}h$
can also be expressed as a boundary term. Recall that we have 
\begin{equation}
\bar{\xi}^{\mu}(\ensuremath{{\cal {G}}}_{\mu\nu})^{(1)}=\bar{\nabla}^{\mu} F_{\mu\nu},
\end{equation}
where $ F^{\nu\mu}$ is antisymmetric in its indices. By
expressing $ F^{\nu\mu}$ and using $\text{\ensuremath{\mathscr{L}}}_{X}h$ instead of $h$,
we can obtain the boundary of the left-hand side. Since we have
\begin{equation}
\bar{\text{\ensuremath{\xi}}}_{\nu}\left(\text{\ensuremath{{\cal {G}}}}^{\nu\mu}\right)^{\left(1\right)}=\bar{\nabla}_{\alpha}\left(\bar{\text{\ensuremath{\xi}}}_{\nu}\bar{\nabla}_{\beta}K^{\mu\alpha\nu\beta}-K^{\mu\beta\nu\alpha}\bar{\nabla}_{\beta}\bar{\text{\ensuremath{\xi}}}_{\nu}\right),
\end{equation}
with the superpotential given as 
\begin{equation}
K^{\mu\alpha\nu\beta}:=\frac{1}{2}\left(\bar{g}^{\alpha\nu}\tilde{h}^{\mu\beta}+\bar{g}^{\mu\beta}\tilde{h}^{\alpha\nu}-\bar{g}^{\alpha\beta}\tilde{h}^{\mu\nu}-\bar{g}^{\mu\nu}\tilde{h}^{\alpha\beta}\right),
\end{equation}
and $\tilde{h}^{\mu\nu}:=h^{\mu\nu}-\frac{1}{2}\bar{g}^{\mu\nu}h$,
we can write 
\begin{equation}
\bar{\xi}_{\nu}\delta_{X}(\ensuremath{{\cal {G}}}^{\mu\nu})^{(2)}=-\bar{\nabla}_{\alpha}\left(\bar{\text{\ensuremath{\xi}}}_{\nu}\bar{\nabla}_{\beta}K^{\mu\alpha\nu\beta}\cdot\text{\ensuremath{\mathscr{L}}}_{X}h-\bar{\nabla}_{\beta}\bar{\text{\ensuremath{\xi}}}_{\nu}\,K^{\mu\beta\nu\alpha}\cdot\text{\ensuremath{\mathscr{L}}}_{X}h\right),
\end{equation}
where $K^{\mu\alpha\nu\beta}$ evaluated at $\text{\ensuremath{\mathscr{L}}}_{X}{h}$ is
\begin{equation}
K^{\mu\alpha\nu\beta}\cdot\text{\ensuremath{\mathscr{L}}}_{X}h=\frac{1}{2}\left(\bar{g}^{\alpha\nu}\text{\ensuremath{\mathscr{L}}}_{X}\tilde{h}^{\mu\beta}+\bar{g}^{\mu\beta}\text{\ensuremath{\mathscr{L}}}_{X}\tilde{h}^{\alpha\nu}-\bar{g}^{\alpha\beta}\text{\ensuremath{\mathscr{L}}}_{X}\tilde{h}^{\mu\nu}-\bar{g}^{\mu\nu}\text{\ensuremath{\mathscr{L}}}_{X}\tilde{h}^{\alpha\beta}\right),
\end{equation}
which altogether shows that under gauge transformations the Taub charge produces a boundary term.

\end{document}